\RequirePackage[l2tabu, orthodox]{nag}
\pdfoutput=1
\documentclass[letterpaper, 10 pt, conference]{ieeeconf}  
\IEEEoverridecommandlockouts  
\usepackage{microtype}
\usepackage[all]{onlyamsmath}
\usepackage{mathtools}
\usepackage{fullpage}
\usepackage{comment}
\usepackage{setspace}
\usepackage{amsmath}
\usepackage{amssymb}
\usepackage{etoolbox}
\usepackage{color,colortbl}
\usepackage{mathtools}
\usepackage{enumerate}
\usepackage{algorithm, algpseudocode}
\usepackage{booktabs}
\usepackage{multirow}
\usepackage[normalem]{ulem}
\usepackage{pgf}
\usepackage{tikz,pgfplots}
\usetikzlibrary{trees,decorations,arrows,arrows.meta,automata,shadows,positioning,plotmarks,backgrounds,shapes,shapes.misc}
\usetikzlibrary{calc,matrix,fit,petri,decorations.pathmorphing,patterns}
\usetikzlibrary{decorations.pathreplacing,decorations.markings,shapes.geometric,calc}
\usepackage{paralist}
\usepackage{stmaryrd}
\usepackage{xspace}
\usepackage{graphicx}
\usepackage{float}
\usepackage[hang, tight]{subfigure}
\usepackage[utf8]{inputenc} 
\usepackage[english]{babel} 
\usepackage{csquotes} 
\usepackage{epstopdf}
\usepackage{array}
\usepackage{xifthen}
\usepackage{url}
\usepackage{relsize}

\newcommand{\REFalg}[1]{Alg.~\ref{#1}}

\newcommand{\REFsec}[1]{Sec.~\ref{#1}}

\newtheorem{definition}{Definition}

\newcommand{\setmap}{\rightrightarrows}
\newcommand{\set}[1]{\{#1\}}

\newcommand{\dom}[1]{\mathrm{dom}(#1)}
\newcommand{\card}[1]{\mathrm{card}\,#1}

\newcommand{\Sys}{\Sigma}
\newcommand{\Xc}{\ensuremath{X}}
\newcommand{\Uc}{\ensuremath{U}}
\newcommand{\fc}{\ensuremath{f}}

\newcommand{\xc}{\ensuremath{x}}

\newcommand{\SysT}{\Sys}
\newcommand{\ft}{\ensuremath{f}}
\newcommand{\ftc}{\ensuremath{\ft^C}}
\newcommand{\traj}{\rho}

\newcommand{\Spec}{\Phi}

\newcommand{\Cfamily}{\mathcal{C}}

\newcommand{\Prob}{\mathcal{P}}
\newcommand{\ProbA}{\widehat{\Prob}}

\newcommand{\Abs}{\Delta}
\newcommand{\Xa}{\ensuremath{Q}}
\newcommand{\Ua}{\ensuremath{A}}
\newcommand{\fa}{\delta}
\newcommand{\fanor}{\fa^{\mathrm{nor}}}
\newcommand{\fadist}{\fa^{\mathrm{dist}}}
\newcommand{\faC}{\fa^{\widehat{C}}}

\newcommand{\xa}{\ensuremath{q}}
\newcommand{\ua}{\ensuremath{u}}
\newcommand{\proc}{\ensuremath{\mathit{FindAbstraction}}}
\newcommand{\procRisk}{\ensuremath{\mathit{FindRiskAwareAbstraction}}}

\newcommand{\Wnor}{\ensuremath{W_{\mathrm{normal}}}}
\newcommand{\Whi}{\ensuremath{W_{\mathrm{high}}}}

\newcommand{\NSpikes}{\ensuremath{\mathit{NumSpikes}}}
\newcommand{\risk}{\mathrm{r}_\Prob}

\algnewcommand\algorithmicinput{\textbf{Input:}}
\algnewcommand\INPUT{\item[\algorithmicinput]}
\algnewcommand\algorithmicparameters{\textbf{Parameters:}}
\algnewcommand\PARAMETERS{\item[\algorithmicparameters]}
\algnewcommand\algorithmicoutput{\textbf{Output:}}
\algnewcommand\OUTPUT{\item[\algorithmicoutput]}
\algdef{SE}[DOWHILE]{Do}{doWhile}{\algorithmicdo}[1]{\algorithmicwhile\ #1}

\title{
Resilient Abstraction-Based Controller Design
}

\author{
Stanly Samuel,
Kaushik Mallik, 
Anne-Kathrin Schmuck, and
Daniel Neider	
\thanks{
S. Samuel is with Indian Institute of Science (IISc), Bengaluru, India. Email: {\tt\small stanly@iisc.ac.in}.
}
\thanks{
K. Mallik, D. Neider, and A.-K. Schmuck are with Max Planck Institute for Software Systems (MPI-SWS), Kaiserslautern, Germany. Email: {\tt\small \{kmallik, neider, akschmuck\}@mpi-sws.org}.
}
}

\begin{document}

\maketitle
\thispagestyle{empty}

\begin{abstract}
 We consider the computation of resilient controllers for perturbed non-linear dynamical systems w.r.t. linear-time temporal logic specifications. We address this problem through the paradigm of Abstraction-Based Controller Design (ABCD) where a finite state abstraction of the perturbed system dynamics is constructed and utilized for controller synthesis. In this context, our contribution is twofold: (I) We construct abstractions which  model the impact of  occasional high disturbance spikes on the system via so called disturbance edges. (II) We show that the application of resilient reactive synthesis techniques to these abstract models results in closed loop systems which are optimally resilient to these occasional high disturbance spikes. We have implemented this resilient ABCD workflow on top of SCOTS and showcase our method through multiple robot planning examples.
\end{abstract}


\section{Introduction}
\label{sec:intro}

With the advent of digital controllers being increasingly used to control safety-critical cyber-physical systems (CPS), there is a growing need to provide formal correctness guarantees of such controllers. A recent approach to achieve this goal is the so-called \emph{Abstraction-Based Controller Design} (ABCD) \cite{belta2017formal,tabuada2009verification}.
ABCD is usually performed in three-steps.  
First, a finite state abstraction is computed from a given non-linear continuous dynamical system by discretizing the state and input space.
Second, given this abstraction and a linear-time temporal logic specification, a discrete control strategy is synthesized. 
Finally, the discrete control strategy is refined to a continuous controller for the given system, which serves as the output of the procedure.
ABCD has been implemented in various tools for a variety of classes of systems and specifications, with numerous improvements over the basic procedure (e.g.,~ \cite{cosyma,rungger2016scots,mascot,pFaces,ROCS}).

An important feature of controllers for CPS is their robustness against modelling uncertainties and unforeseen operating conditions. The ABCD workflow therefore has a build-in robutstification step; given a uniform upper bound $W$ on the uncertainty of continuous trajectories, the abstract transition system over-approximates all possible trajectories of the original system. 

This approach, however, has one important disadvantage. Consider for example a mobile robot serving coffee in an office building. As it needs to get to the kitchen to get coffee, it sometimes passes through water spilled on the floor which results in an imprecise motion actuation.

In this case, increasing $W$ to contain all possible disturbances caused by spilled water would result in a very conservative abstract model which
renders the resulting abstract controller synthesis problem (e.g., always eventually serving a requested coffee) unrealizable. A more optimistic choice of $W$ allows to design a controller but sacrifices rigorous correctness guarantees --  if water is spilled, the specification might be violated (e.g., the robot might bump into a door frame). Intuitively we however know that it is very unlikely that water is spilled \emph{everywhere} and \emph{all the time}, which would indeed make the specification unrealizable. Further, knowing that there might be spilled water, we would like the robot motion controller to be \enquote{risk-aware} (i.e., to avoid going close to the door frames). 

\smallskip
\noindent\textbf{Contribution.}
In our work, we automatically synthesize a controller which is correct w.r.t.\ a nominal disturbance $\Wnor$ (e.g., no water spilled on the floor), and in addition, is \enquote{risk-minimizing} w.r.t.\ larger disturbance spikes in $\Whi\supset \Wnor$ that may occur any time. 

Formally, we build on top of the basic three step procedure of ABCD. 
First, we obtain a \emph{risk-aware} finite state abstraction for the given continuous dynamical system by adapting the framework of feedback-refinement relations \cite{ReissigWeberRungger_2017_FRR}.
Second, we compute a maximally-resilient control strategy for this abstraction by an adaptation of the two-player game algorithm to compute \emph{Optimally Resilient Strategies} as defined by Neider et al.~\cite{dallal2016synthesis,DBLP:journals/acta/NeiderWZ20}. 
Finally, the control strategy so obtained can be refined to a continuous controller for the original system, while retaining the correctness guarantees and the resilience of the states.
This three step procedure is what we call the \emph{Resilient Abstraction-Based Controller Design}.

\smallskip
\noindent\textbf{Related Work.}
Ensuring robustness of discrete, \emph{event-based} control strategies is an active field of research.  
Within \emph{resilient ABCD}, we employ the method introduced in \cite{dallal2016synthesis,DBLP:journals/acta/NeiderWZ20}, 
where \emph{disturbance edges} are utilized to formalize resilience and are assumed to be given. Unfortunately, a good disturbance model is not always easy to obtain. 
Our \emph{resilient ABCD} method \emph{automatically} injects disturbance edges into the abstract model. 

Another approach to robust reactive synthesis considers 
particular specifications of the form $A\rightarrow G$; in every environment that fulfills the assumption $A$ the controller needs to enforce the property $G$. In this context robustness is for example understood w.r.t.\ assumption violations \cite{bloem2014handle,EhlersTopcu_14, Hagihara_16}, 
hidden or missing inputs \cite{ Bloem_19} or unexpected jumps in the game graph
\cite{DLMurray_ICCPS17}. 
While the abstraction generated within our \emph{resilient ABCD} method can be understood as a particular safety assumption $A$ for the given synthesis problem interpreted as $G$, the assumption violations we consider are already explicitly modeled by \emph{automatically generated} disturbance edges. This avoids analyzing assumptions for possible faults and allows for a more fine-grained analysis assigning resilience values to states rather than whole systems.
While lifting the synthesis of robust controllers to the abstract domain, \emph{resilient ABCD} still retains the intuitive local nature of \emph{robustness} of continuous systems by injecting disturbance edges only locally. This is closely related to work in robust CPS design, where continuous and discrete metrics are imposed to construct abstractions for robust controller synthesis \cite{RunggerTabuada_TAC16, tabuada2014towards,LiuOzay_16}. In this context, robustness ensures that the controlled system remains \enquote{close} to a chosen execution path despite disturbances.
In some cases, a similar sort of robustness is achieved with the help of online monitoring of the system and updating the controller if unexpected events occur \cite{wongpiromsarn2012receding}.
In contrast, \emph{resilient ABCD} exploits the existence of multiple different control strategies and picks the most resilient among them. This optimizes resilience over the infinite time horizon of the system's execution.

An alternate approach to ours would be to assign probability values to disturbances: The ones in $\Whi\setminus \Wnor$ occur with low probability, and the ones in $\Wnor$ occur with high probability.
Then an optimal controller that maximizes the probability of satisfaction of the given specification \cite{dutreix2020abstraction, majumdar2020symbolic, svorevnova2017temporal} would perhaps behave similarly to the optimally resilient controller in our setup.
In contrast to our approach, there are two drawbacks of the probabilistic treatment of unexpected disturbances: 
First, an explicit probabilistic model of the disturbance is required, which is often difficult to obtain.
Second, just by knowing the optimal satisfaction probability of the specification from a given state, one cannot tell whether an absence of larger disturbance would surely ensure the satisfaction of the specification.

For particular classes of continuous-time systems and temporal logic specifications, controllers can be designed without explicitly constructing an abstraction by state-space gridding. In this line of work, robustness requirements can be specified by  signal temporal logic (STL) formulas and enforced through \enquote{classical} optimal robust controller synthesis methods \cite{sadraddini2015robust,lindemann2019robust,mehdipour2019arithmetic}.
 
Our \emph{resilient ABCD} method is orthogonal to this line of work, as we \emph{lift} the treatment of robustness to the abstract domain. This allows us to handle more general specification classes, namely full linear temporal logic (LTL), and arbitrary non-linear continuous dynamics. 
Further, \emph{resilient ABCD} allows to incorporate other discrete disturbances s.a.\ lossy channels or faulty event models \cite{girault2009automating,YuJiang_review_15}.

\section{Preliminaries on Control Systems}

\smallskip
\noindent\textbf{Notation.}
We introduce an ordinal notation in the style of von Neumann.
The nonnegative integers are inductively defined as $0 = \emptyset$ and $n+1 = n \cup \{ n \}$.
Consequently, the first limit ordinal is the set of non-negative integers $\omega = \{ 0, 1, \ldots \}$ and the next two ordinals are $\omega+1$ and $\omega+2$.
It is easy to see that these ordinals are ordered by set inclusion. 
For any given set $S$, we use the notation $S^\omega$ to represent the set of all infinite sequences that can be formed by using the elements of the set $S$.

\smallskip
\noindent\textbf{Sampled-time Control System.}
A \emph{sampled-time control system} is a tuple $\SysT = (\Xc,\Uc,\Wnor,\ft)$, which consists of a state space $\Xc= \mathbb{R}^n$, a non-empty input space $\Uc\subseteq \mathbb{R}^m$, 
a bounded disturbance set $\Wnor\subset \mathbb{R}^n$,
and a transition function $\ft:\Xc\times \Uc\times\mathbb{R}^n \rightarrow \Xc$.

\smallskip
\noindent\textbf{Control Specifications.}
Let $\mathit{Win}\subseteq \Xc^\omega$ be a given control specification.
We consider two different control specifications: safety and parity (the latter being the canonical representation of temporal logic specifications).
For safety the set $\mathit{Win}$ is defined by $\mathit{Safety}(W) = \{ x_0 x_1 \ldots \in \Xc^\omega \mid x_i \notin W \text{ for all } i \in \omega \}$ for a given set $W \subseteq \Xc$ of unsafe states and requires that the systems state be always outside the set $W$.
For parity conditions the set $\mathit{Win}$ is defined by $\mathit{Parity}(\Phi) \coloneqq \{ x_0 x_1 \ldots \in \Xc^\omega \mid \limsup\set{\Phi(x_0), \Phi(x_1), \ldots} \text{ is even} \}$ for a function $\Phi \colon \Xc \to \omega$ assigning a priority to states, and requires that the maximum priority seen by the system infinitely often is even.
As per convention, we will use the term ``\emph{color}'' in place of ``priority'' while using parity conditions.
For notational convenience, we use $\Phi$ and $W$ in place of $\mathit{Parity}(\Phi)$ and $\mathit{Safety}(W)$, respectively.

Both parity and safety conditions permit memoryless (or positional) control strategies, which only depend on the current state.
This allows us to restrict ourselves to static state feedback controllers, introduced next.

\smallskip
\noindent\textbf{Controller and Closed-loop.}
A state-feedback controller, or simply a controller, for the control system $\SysT=(\Xc,\Uc,\Wnor,\ft)$ is a partial function $C:\Xc\rightarrow \Uc$.
We denote the closed loop formed by connecting $C$ to $\SysT$ in feedback as $\SysT\parallel C = (\dom{C},\mathbb{R}^n,\ftc)$, where $\dom{C} \subseteq \Xc$ is the domain of the controller $C$, the transition function $\ftc:\dom{C} \times \mathbb{R}^n \rightarrow\Xc$ is obtained from $\ft$ by using the rule $\ftc:(\xc,w)\mapsto \ft(\xc,C(\xc),w)$.

A closed-loop \emph{trajectory} of $\SysT\parallel C$ starting at a state $\xc_0\in \dom{C}$, exposed to disturbances from the set $W\subseteq \mathbb{R}^n$, is an infinite sequence $\traj^W(\xc_0)= (\xc_0,w_0)(\xc_1,w_1)\ldots$ s.t.\ for all $i\in \omega$, $w_i\in W$, and $\xc_{i+1}=\ftc(\xc_i,w_i)$.
The trajectory $\traj^W(\xc_0)$ is said to satisfy a given specification $\mathit{Win}$, denoted by $\traj^W(\xc_0)\vDash\mathit{Win}$, if the sequence $\xc_0\xc_1\ldots$ satisfies $\mathit{Win}$. Otherwise $\traj^W(\xc_0)$ violates $\mathit{Win}$, denoted by $\traj^W(\xc_0)\nvDash\mathit{Win}$.

A controller $C$ is called \emph{sound} if for all $\xc\in \dom{C}$, and for every trajectory $\traj^{\Wnor}(\xc)$ (for every possible disturbance sequence from $\Wnor$), $\traj^{\Wnor}(\xc)\vDash\mathit{Win}$.
A sound controller $C$ is called \emph{maximal} if there is no other sound controller $C'$ s.t.\ $\dom{C'}\supset \dom{C}$.
We write $\Cfamily^{\SysT,\mathit{Win}}$ for the set of all sound and maximal controllers for a system $\SysT$ and a specification $\mathit{Win}$.

\section{Problem Statement}\label{sec:problem}

We define a \emph{risk-aware controller synthesis problem} using the tuple $\Prob=(\SysT,\Whi,\Spec)$, where $\SysT=(\Xc,\Uc,\Wnor,\fc)$ is a sampled-time control system that is normally exposed to disturbances from $\Wnor$, the set $\Whi\supset \Wnor$ is a set of higher disturbance spikes that $\Sys$ is only occasionally exposed to, and $\Spec$ is a parity specification over the state space $\Xc$.

For a risk-aware controller synthesis problem, we focus on the trajectories $\traj^{\Whi}(\xc_0)$ with disturbance values from the set $\Whi$, rather than $\traj^{\Wnor}(\xc_0)$.
Note that in practice, we interpret disturbances from the set $\Whi\setminus \Wnor$ as rare events.
Given any trajectory $\traj^{\Whi}(\xc_0)=(\xc_0,w_0)(\xc_1,w_1)\ldots$, define the operator $\NSpikes: \traj^{\Whi}(\xc_0)\mapsto \card \set{i \in \omega \mid w_i\in \Whi\setminus \Wnor}$, where $\card{S}$ is the cardinality of a given set $S$, and it is understood that $\Wnor$ is clear from the context.
Intuitively, $\NSpikes(\traj^{\Whi}(\xc_0))$ returns the number of times a disturbance outside $\Wnor$ (but in $\Whi$) occurred in the trajectory $\traj^{\Whi}(\xc_0)$.
A closed-loop trajectory $\traj^{\Whi}(\xc_0)$ is called \emph{spike-free} if $\NSpikes(\traj^{\Whi}(\xc_0))=0$.

Let $C\in \Cfamily^{\SysT,\Spec}$ be a controller, and $\alpha \in \omega + 2$.
We say that $C$ is \emph{$\alpha$-resilient} from a state $\xc_0 \in \Xc$ of $\SysT$, if every closed-loop trajectory $\traj^{\Whi}(\xc_0)$ of $\SysT\parallel C$ that starts in $\xc_0$ and satisfies $\NSpikes(\traj^{\Whi}(\xc_0)) < \alpha$ also satisfies the specification $\Spec$.
This means that a $k$-resilient controller with $k \in \omega$ satisfies $\Spec$ even under at most $k-1$ high disturbance spikes, an $\omega$-resilient controller satisfies $\Spec$ even under any finite number of high disturbance spikes, and an $(\omega + 1)$-resilient controller satisfies $\Spec$ even under infinitely many high disturbance spikes.

Next, we define the \emph{resilience} of a state $\xc \in \Xc$ to be
\begin{multline*}
	r_{\Prob}(\xc) = \sup \{ \alpha \in \omega + 2 \mid \text{there exists an} \\
	\text{$\alpha$-resilient controller from $\xc$} \}.
\end{multline*}
Note that $r_{\Prob}(\xc) > 0$ if and only if $\xc \in \dom{C}$, because any controller is $1$-resilient from $\xc$ if and only if $\xc\in \dom{C}$. 

We call a controller $C^*\in \Cfamily^{\SysT,\Spec}$ \emph{optimally resilient}, if it is $r_{\Prob}(\xc)$-resilient from every state $\xc \in \Xc$.

In this paper, we address the following  problem: 

\emph{Given a risk-aware controller synthesis problem $\Prob$, find a procedure that will automatically synthesize an \emph{approximately} optimally resilient controller $C$}.

Following the paradigm of ABCD, we address this problem in three steps. We
\begin{inparaenum}[(i)]
 \item reduce the \emph{risk-aware controller synthesis problem} $\Prob$ into an \emph{abstract risk-aware controller synthesis problem} $\ProbA$ (see Sec.~\ref{sec:compute_abs}),
  \item we synthesize an \emph{optimally resilient controller}  $\widehat{C}^*$ for the problem $\ProbA$ (see Sec.~\ref{sec:synth}), and
 \item we refine  $\widehat{C}^*$ into an approximately optimally resilient controller $C$ for $\Prob$ which approaches the optimal solution $C^*$ if the grid size for the state and input grid goes to zero.
\end{inparaenum}

It is unknown whether the optimally resilient controller $C^*$ for the problem $\mathcal{P}$ always exists.
However, we show that an optimally resilient controller for the abstract problem $\widehat{\mathcal{P}}$ exists. We generate the controller $C$ by refining the optimally resilient controller $\widehat{C}^*$ for $\widehat{\mathcal{P}}$.

\section{Risk-aware Abstractions}
\label{sec:compute_abs}

\subsection{Preliminaries}\label{sec:prelims_abcs}

The method that we primarily build upon is Abstraction-Based Control Design (ABCD).
In the following, we briefly recapitulate one of the several available ABCD techniques.

\smallskip
\noindent\textbf{Finite State Transition System.}
A \emph{finite state transition system} is a tuple $\Delta= (\Xa, \Ua, \fa)$ that consists of a finite set of states $\Xa$, a finite set of control actions $\Ua$, and a set-valued transition map $\fa:\Xa\times \Ua\setmap \Xa$.

\smallskip
\noindent\textbf{Finite State Abstraction of Control Systems.}
A finite state transition system $\Delta$ is called a \emph{finite state abstraction}, or simply an abstraction, of a given control system $\Sys$ if a certain relation holds between the transitions of $\SysT$ and the transitions of $\Delta$.
Depending on the controller synthesis problem at hand, there are several such relations available in the literature.
The one that we use in our work is the \emph{feedback refinement relation} (FRR) \cite{ReissigWeberRungger_2017_FRR}.

An FRR is a relation $R\subseteq \Xc\times \Xa$ between $\SysT$ and a finite-state transition system $\Delta$, written as $\SysT \preccurlyeq_R \Delta$, so that for every $(\xc,\xa)\in R$, the set of allowed control inputs (element of $\Ua$) from $\xa$ is a subset of the set of allowed control inputs (element of $\Uc$) from $\xc$, and moreover when the same allowed control input is applied to both $\xa$ and $\xc$, the image of the set of possible successors of $\xc$ under $R$ is contained in the set of successors of $\xa$.

Within this paper we assume that there is an algorithm, called $\proc$, which takes as input a given control system $\Sys$ and a given set of additional tuning parameters $P$ (like the state space discretization and the control space discretization), and outputs an abstract finite state transition system $\Delta$ and an associated FRR $R$ s.t.\ $\SysT\preccurlyeq_R \Delta$.
For the actual implementation of $\proc(\Sys,P)$, we refer the reader to \cite{ReissigWeberRungger_2017_FRR}.

Without going into the details of how the parameter set $P$ influences the outcome of $\proc$, we would like to point out one property that we expect to hold.
Let $\SysT=(\Xc,\Uc,\Wnor,\fc)$ and $\SysT'=(\Xc,\Uc,\Wnor',\fc')$ be two different control systems with same state and control input spaces.
Suppose $\Abs=(\Xa,\Ua,\fa)$ and $\Abs'=(\Xa',\Ua',\fa')$ be two transition systems computed using $\proc$ using the same parameter set $P$ (i.e.,\ $\proc(\SysT,P)=\Abs$ and $\proc(\SysT',P)=\Abs'$).
Then there are one-to-one correspondences between $\Xa$ and $\Xa'$, and between $\Ua$ and $\Ua'$.
Moreover, there is an FRR $R$ so that both $\SysT\preccurlyeq_R \Abs$ and $\SysT'\preccurlyeq_R \Abs'$ hold.
We will abuse this property, and will use the same state space and input space for both $\Abs$ and $\Abs'$.

\subsection{Risk-aware Abstractions}

To take into account the effect of occasional occurrences of disturbance spikes in the control system $\SysT$, we introduce a special finite state abstract transition system, called the \emph{bimodal transition system}, with two distinct transition relations.
A bimodal transition system is a tuple $(\Xa,\Ua,\fanor,\fadist)$, where $\Xa$ is a finite set of \emph{abstract states}, $\Ua$ is the finite set of control actions, $\fanor:\Xa\times \Ua\setmap \Xa$ is a set-valued map representing the \emph{normal transitions}, and $\fadist:\Xa\times \Ua\setmap \Xa$ is another set-valued map representing a set of \emph{disturbance transitions}.

We use the notation $\Ua(\xa)$ to denote the set of control inputs allowed in the state $\xa\in \Xa$ (i.e.,\ $\Ua(\xa):=\set{u\in\Ua\mid \fanor(\xa,u)\neq \emptyset}$).

\begin{definition}\label{def:risk-aware abstraction}
	Let $(\SysT,\Whi,\Spec)$ be a risk-aware controller synthesis problem.
	A bimodal transition system $\Gamma=(\Xa,\Ua,\fanor,\fadist)$ is called a \emph{risk-aware abstraction} of $\SysT$ if there exists a relation $R\subset \Xc\times \Xa$ s.t.\ the following conditions are satisfied:
	\begin{itemize}
		\item $\SysT \preccurlyeq_R (\Xa,\Ua,\fanor)$, and
		\item $ \text{for all }(\xc,\xa)\in R$ and for all $u\in \Ua(\xa)$, the following holds: $ \cup_{w\in \Whi}R(\ft(\xc,u,w))\subseteq \fanor(\xa,u)\cup \fadist(\xa,u)$.
	\end{itemize}
\end{definition}

\begin{algorithm}
	\caption{$\procRisk$}
	\label{alg:compute risk-aware abstraction}
	\begin{algorithmic}[1]
		\INPUT $(\SysT,\Whi,\Spec)$, some parameter set $P$ for $\proc$
		\OUTPUT $\Gamma=(\Xa,\Ua,\fanor,\fadist)$ 
		\State Compute $\Abs^{\mathrm{nor}}=(\Xa,\Ua,\fanor)\gets\proc(\SysT,P)$
		\State Compute $\Abs^{\mathrm{high}}=(\Xa,\Ua,\fa^{\mathrm{high}})\gets\proc\left((\Xc,\Uc,\Whi,\fc),P\right)$
		\State For all $\xa\in \Xa$ and $\ua\in \Ua$, define $\fadist:(\xa,\ua)\mapsto \fa^{\mathrm{high}}(\xa,\ua)\setminus \fanor(\xa,\ua)$
		\State \Return $\Gamma=(\Xa,\Ua,\fanor,\fadist)$
	\end{algorithmic}
\end{algorithm}

\REFalg{alg:compute risk-aware abstraction} computes a risk-aware abstraction for a given risk-aware controller synthesis problem and a given parameter set $P$.
\REFalg{alg:compute risk-aware abstraction} can be implemented symbolically.


\section{Synthesis of Optimally Resilient Controllers}\label{sec:synth}

In the following, we present our algorithm to synthesize the optimally resilient controller for risk-aware abstractions of sampled-time control systems.
Our algorithm follows the general ideas of \cite{DBLP:journals/acta/NeiderWZ20}, and assigns resilience values to every abstract state of the risk-aware abstraction.
Intuitively, the resilience of an abstract state corresponds to the maximum number of disturbance spikes that the \emph{abstract closed-loop}---formed by connecting the synthesized controller with the abstract system in feedback---can tolerate while still satisfying its specification.
We are interested to synthesize controllers which \emph{maximize} the resilience of each abstract state.
Before we describe our algorithm we introduce required definitions and notation.

\subsection{Preliminaries}

Let $\Prob=(\SysT,\Whi,\Spec)$ be a given risk-aware controller synthesis problem, $\Gamma=(\Xa,\Ua,\fanor,\fadist)$ a risk-aware abstraction of $\SysT$, and $R$ the associated FRR between $\SysT$ and $(\Xa,\Ua,\fanor)$.
We assume that the parity specification $\Spec$ is so given and $\Gamma$ is so obtained that for every abstract state $q\in \Xa$, every pair of associated system states $x_1,x_2 \in \Xc$ with $(x_1,q),(x_2,q)\in R$ are assigned the same color by $\Spec$ (i.e.,\ $\Spec(x_1)=\Spec(x_2)$).
This allows us to lift the parity specification $\mathit{Parity}(\Spec)$ in a well-defined way to the abstract state space as $\mathit{Pairty}(\widehat{\Spec})\subseteq \Xa^\omega$, s.t.\ for all $q\in \Xa$ and for all $x\in \Xc$ with $(x,q)\in R$, $\widehat{\Spec}:q\mapsto \Spec(x)$.
We denote the \emph{abstract risk-aware controller synthesis problem}, or abstract synthesis problem in short, by $\ProbA=(\Gamma,\widehat{\Spec})$.

We export the concepts of controllers, closed-loops, closed-loop trajectories etc.\ from the domain of sampled-time control systems to the domain of risk-aware abstractions in the natural way.
An abstract controller $\widehat{C}$ is a partial function $\widehat{C}:\Xa\rightarrow \Ua$.
The abstract closed-loop is obtained by connecting $\widehat{C}$ in feedback with $\Gamma$, and is defined using the tuple $\Gamma\parallel \widehat{C}=(\dom{\widehat{C}}, \faC)$, where $\faC:\dom{\widehat{C}}\setmap \Xa$ s.t.\ $\faC:q\mapsto \fanor(q,\widehat{C}(q))\cup \fadist(q,\widehat{C}(q))$.
The notion of trajectory, the satisfaction of specification, $\NSpikes$, spike-free trajectories, and resilience are naturally adapted for the system $\Gamma$.
It should be noted, however, that $\NSpikes$ and resilience are now computed by \emph{counting the number of $\fadist$-transitions} appearing in any given abstract closed-loop trace, as opposed to counting the number of disturbances appearing from the set $\Whi\setminus\Wnor$ in a given sampled-time closed-loop trace.

Like in the case of controller synthesis problem $\Prob$, we require our abstract controllers to be sound and maximal.
Such a controller is called a \emph{winning controller} w.r.t.\ a given abstraction specification $\mathit{Win}\subseteq \Xa^\omega$, and the respective controller domain is called the winning region $\mathcal W(\ProbA)$.
Let $\mathcal{C}^{\Gamma,\mathit{Win}}$ be the set of winning abstract controllers for the abstract synthesis problem $\ProbA=(\Gamma,\mathit{Win})$.
We assume that we have access to a solver for the sound and maximal abstract controllers for the (spike-free) safety and parity specifications, which serves as a black-box method in our synthesis routine (for the actual implementation, one can use any of the available methods from the literature \cite{van2018oink}).
Such a solver takes a controller synthesis problem $\ProbA$ with safety, parity, or a conjunction of safety and parity specification as input, and outputs the winning region $\mathcal W(\ProbA)$ (as well as the complement $\overline{\mathcal W}(\ProbA)$) together with a (uniform) abstract controller $\widehat{C}$ that is winning for every abstract state $\xa \in \mathcal W(\ProbA)$.


We define the resilience of abstract states $r_{\ProbA}(\xa)$ and the optimally resilient abstract controller $\widehat{C}^*$ analogously to the continuous system defined in \REFsec{sec:problem}.
Recall that a $k$-resilient control strategy with $k \in \omega$ is winning even under at most $k-1$ disturbance spikes, an $\omega$-resilient strategy is winning even under any finite number of disturbance spikes, and an $(\omega + 1)$-resilient strategy is winning even under infinitely many disturbance spikes.

\subsection{Computing Optimally Resilient Strategies}
Following Neider, Weinert, and Zimmermann~\cite{DBLP:journals/acta/NeiderWZ20}, we first characterize the abstract states of finite resilience.
The remaining abstract states then have either resilience $\omega$ or $\omega +1$, and we show how to distinguish between them.
Finally, we describe how to derive an optimally resilient abstract controller based on the computed resilient values.

\subsubsection{Finite Resilience}
Starting from the abstract states in $\overline{\mathcal W}(\ProbA)$, which have resilience $0$ by definition, we use two operations to determine the abstract states of finite resilience: the disturbance update and the risk update.
Intuitively, the disturbance update computes the resilience of abstract states for which a disturbance spike (i.e., a transition in $\fadist$) leads to an abstract state whose resilience is already known.
The risk update, on the other hand, determines the resilience of abstract states from which the controller can either not prevent to visit an abstract state with known resilience or it needs to move to such an abstract state in order to avoid losing.

For the remainder, let us fix an abstract synthesis problem $\ProbA = (\Gamma, \widehat{\Phi})$ with risk-aware abstraction $\Gamma=(\Xa,\Ua,\fanor,\fadist)$.
Following Neider, Weinert, and Zimmermann~\cite{DBLP:journals/acta/NeiderWZ20}, we define the disturbance and risk updates as updates on partial mappings $r \colon \Xa \to \omega$, which are called \emph{rankings}.
Intuitively, a ranking assigns resilience to some of the abstract states.
We denote the domain of $r$ by $\mathrm{dom}(r)$ and the image of $r$ by $\mathrm{im}(r)$.

Due to the different problem setup, we now deviate from Neider, Weinert, and Zimmermann's original algorithm in that we perform disturbance and risk updates not on the same risk-aware abstraction but on a sequence $(\Gamma_i)_{i=1, 2, \ldots}$ of abstractions.
We obtain $\Gamma_{i+1}$ from $\Gamma_i$ using an operation we call \emph{strategy pruning}, which removes certain transitions from $\Gamma_i$ that are no longer relevant for the computation of resilience values.

Strategy pruning is inter-weaved with the disturbance and risk updates as shown in Algorithm~\ref{alg:computing-finite-resilience}.
Our algorithm starts with the \emph{initial ranking} that assign the value $0$ to all $\xa \in \overline{\mathcal W}(\Gamma)$ and is otherwise undefined.
Then, it computes the disturbance update on $\Gamma_i$, applies strategy pruning to obtain $\Gamma_{i+1}$, and finally computes the risk update on $\Gamma_{i+1}$.
This process repeats until a fixed point is reached (i.e., $r_i = r_{i-1}$), at which point the algorithm returns the final ranking $r^\ast = r_i$.
The ranking $r^\ast$ then maps an abstract state to its resilience.

\begin{algorithm}
	\begin{algorithmic}[1]
		\INPUT Risk-aware abstraction $\Gamma$
		
		\State Compute $\overline{\mathcal W}(G)$
		\State Initialize an initial ranking $r_0$ with $r_0(\xa) = 0$ for all $\xa \in \overline{\mathcal W}(G)$ and \emph{undefined} otherwise
		\State $i \gets 0$ and $\Gamma_0 \gets \Gamma$
		
		\Repeat
			\State $r' \gets \mathit{disturbance\textunderscore{}upd}(r_i, \Gamma_i)$
			\State $\Gamma_{i+1} \gets \mathit{strategy\textunderscore{}pruning}(r_i,\Gamma_i)$
			\State $r_{i+1} \gets \mathit{risk\textunderscore{}upd}(r', \Gamma_{i+1})$
			\State $i \gets i + 1$
		\Until{$r_i = r_{i-1}$}

		\State \Return $r^\ast = r_i$

	\end{algorithmic}
	\caption{Determining finite resilience} \label{alg:computing-finite-resilience}
\end{algorithm}

In the remainder of this section, we describe all three operations in detail.

\paragraph{Disturbance Update}
As mentioned before, the intuition behind the disturbance update, which we denote by $\mathit{disturbance\textunderscore{}upd}$, is to compute the resilience of abstract states for which a disturbance spike leads to an abstract state whose resilience is already known.
It takes two inputs: a ranking $r$ and a risk-aware abstraction $\Gamma = (\Xa, \Ua, \fanor, \fadist)$.

In the first step, the disturbance update computes for each $\xa \in \Xa$ a set $S_\xa\subseteq \Xa$ which satisfies the condition given below: 
\begin{align*}
	\xa'\in S_\xa \Leftrightarrow {} & [\text{for all $\ua \in \Ua$,} \\
	& \fanor(\xa, \ua) \neq \emptyset \text{ and } \fanor(\xa, \ua) \cap \mathrm{dom}(r) = \emptyset \\
	& \text{imply } \fadist(\xa, \ua) \cap \mathrm{dom}(r) \neq \emptyset] \\
	& \text{and } \\
	& [\text{there exists $\ua \in \Ua$,} \\
	& \fanor(\xa, \ua) \neq \emptyset \text{ and } \fanor(\xa, \ua) \cap \mathrm{dom}(r) = \emptyset \\
	& \text{and } \xa' \in \fadist(\xa, \ua) \text{ and } \xa' \in \mathrm{dom}(r)].
\end{align*} 
Then, it returns a new ranking $r'$ with
\[ r'(\xa) = \min \bigl\{ \{ r(\xa) \} \cup \{ r(\xa') + 1 \mid \xa' \in S_\xa \} \bigr \} \]
for each $\xa \in \Xa$, where $\{ r(\xa) \} = \emptyset$ if $\xa \notin \mathrm{dom}(r)$, and $\min{\emptyset}$ is undefined. 

\paragraph{Strategy Pruning}
The strategy pruning step, which we denote by $\mathit{strategy\textunderscore{}pruning}$, removes transitions from a risk-aware abstraction that are no longer relevant for the computation of resilience values.
It takes two inputs: a ranking $r$ and a  risk-aware abstraction $\Gamma = (\Xa, \Ua, \fanor, \fadist)$.

Based on the transition function $\fanor$ and $\fadist$, strategy pruning first computes two new transition functions ${\fanor}'$ and ${\fadist}'$ where
\begin{align*}
	{\fanor}'(\xa, \ua)  & = \begin{cases} \emptyset & \text{if $F$ is true; and} \\ \fanor(\xa, \ua)  & \text{otherwise} \end{cases}
	\intertext{for all $\xa \in \Xa$ and $\ua \in \Ua$ as well as}
	{\fadist}'(\xa, \ua)  & = \begin{cases} \emptyset & \text{if $F$ is true; and} \\ \fadist(\xa, \ua)  & \text{otherwise} \end{cases}
\end{align*}
for all $\xa \in \Xa$ and $\ua \in \Ua$, and where $F$ is true if and only if $\fanor(\xa, \ua) \cap \mathrm{dom}(r) \neq \emptyset$ or $\fadist(\xa, \ua) \cap \mathrm{dom}(r) \neq \emptyset$.
Then, it returns the new risk-aware abstraction $\Gamma' = (\Xa, \Ua, {\fanor}', {\fadist}')$.

\paragraph{Risk Update}
The risk update, which we denote by $\mathit{risk\textunderscore{}upd}$, determines the resilience of abstract states from which the controller can either not prevent to visit an abstract state with known resilience or it needs to move to such an abstract state in order to avoid losing.
Like the disturbance update, it takes two inputs: a ranking $r$ and a risk-aware abstraction $\Gamma = (\Xa, \Ua, \fanor, \fadist)$.

For each $k \in \mathrm{im}(r)$, the risk update computes the set
\begin{multline*}
	B_k = \overline{\mathcal W} \bigl(\Gamma, \mathit{Parity}(\Phi) \cap {} \\
	\mathit{Safety}( \{x \in \mathrm{dom}(r) \mid r(x) \leq k \}) \bigr).
\end{multline*}
As described above, this can be done by first solving the safety game and then the parity game on the resulting game.
As a byproduct, we obtain in iteration $i$ a controller that (i) never visits a state with resilience less than $i$ and (ii) is winning if no disturbance spike occurs.
Once the fixed point is reached, the final controller can even tolerate an arbitrary number of resilience spikes.

Then, the risk update returns the new ranking $r'$ with
\[ r'(\xa) = \min{\{ k \mid \xa \in B_k \}}, \]
for every $\xa \in \Xa$, where again $\min{\emptyset}$ is undefined.
This concludes the computation of finite resiliences.

\subsubsection{Distinguishing Between Resilience $\omega$ and $\omega+1$}
It is left to distinguish between abstract states of resilience $\omega$ and $\omega + 1$.
Recall that abstract states of resilience $\omega + 1$ are those from which the controller can satisfy the specification even if infinitely many disturbance spikes occur.
Fortunately, characterizing the abstract states of resilience $\omega + 1$ is straightforward.

We solve the classical controller synthesis problem for the abstraction $(\Xa, \Ua, \fanor\cup \fadist)$ (i.e., the abstraction that always allows for large disturbances), where $\fanor\cup\fadist$ denotes the argument-wise union operation of the set-valued transition functions $\fanor$ and $\fadist$.
In fact, it is  then not hard to verify that the abstract states in $\mathrm{dom}(\widehat{C})$ of the resulting abstract controller $\widehat{C}$ are exactly those of resilience $\omega + 1$, and the controller is $(\omega + 1)$-resilient from these abstract states.
This is due to the fact that $\widehat{C}$ can enforce the specification even under arbitrary many occurrences of high disturbances.

In total, we have identified the abstract states of finite resilience and resilience $\omega + 1$ of a risk-aware abstraction $\Gamma = (\Xa, \Ua, \fanor, \fadist)$.
The remaining abstract states in $\Xa$ must then have resilience $\omega$.
This concludes the computation of the function $r^\ast$, which maps an abstract state to its resilience, and we can  now describe how to extract an optimally resilient controller.

\subsubsection{Extracting Optimally Resilient Strategies}
The extraction of an optimally resilient abstract controller follows very closely the one by Neider, Weinert, and Zimmermann~\cite{DBLP:journals/acta/NeiderWZ20}. 
Intuitively, the controller extraction is the process of stitching together the controllers that were obtained in different iterations of the risk update and the $(\omega+1)$-resilience computation.
The underlying idea is to switch the controller whenever a disturbance spike occurs.
If finitely many disturbance spikes occur, then the optimally resilient controller will settle with an appropriate \mbox{(sub-)}controller that was obtained in the risk update.
If infinitely many disturbance spikes occur, then the optimally resilient controller will settle with the \mbox{(sub-)}controller obtained in the $(\omega+1)$-resilience computation.
In both cases, these (sub-)controllers are winning by construction and, hence, so is the optimally resilient one.

\subsubsection{Controller refinement}
Let $\Prob=(\SysT,\Whi,\Spec)$ be a risk-aware controller synthesis problem.
Let  $\widehat{C}^*$ be  an optimally resilient controller synthesized for the abstract synthesis problem $\widehat{\Prob}=(\Gamma,\widehat{\Spec})$, where $\Gamma$ is a risk-aware abstraction of $\SysT$ with the corresponding FRR $R$.
Following the usual methodology of ABCD, one can define a controller for $\SysT$ as $C:\xc\mapsto \widehat{C}^*(\xa)$ where $\xa\in \Xa$ s.t.\ $(\xc,\xa)\in R$. Here, $C$ is well-defined due to the properties of $R$.
It can be shown that for every $\xc\in \Xc$, $C$ is an $\alpha$-resilient controller for some $\alpha \leq \risk(\xc)$ (i.e.,\ $C$ is a sub-optimal approximation of the optimally resilient controller $C^*$).

\section{Experimental Results}\label{sec:experiments}

We have implemented our \emph{resilient ABCD} approach in the tool RESCOT (REsilient SCOTS).
RESCOT is built on top of the existing ABCD tool called SCOTS \cite{rungger2016scots}, and is available on Bitbucket.\footnote{\url{https://bitbucket.org/stanlyjs/rescot/src/master/}}

We have evaluated RESCOT on various unicycle motion planning problems, with the unicycle dynamics being adapted from \cite{ReissigWeberRungger_2017_FRR} with nominal disturbance set $\Wnor = [-0.05,0.05]\times [-0.05,0.05]\times [0,0]$. In addition, we assume that the velocity at each dimension of the unicycle will occasionally be perturbed (e.g.,\ due to a slippery floor) by disturbances spikes from the set $\Whi=[-d,d]\times [-d,d]\times [0,0]$ for some $d>0.05$. 

We list the computation times and controller sizes for all run experiments in Table~\ref{tab:1}. The experiments were performed on a system with 16 core 3.3 GHz Intel Xeon E5-2667 v2 CPU with 256GB memory. However, it must be noted that parallelism was not exploited. 

\begin{figure}
 	\centering
	 \includegraphics[trim={3.3cm 1.8cm 2.8cm 1.8cm}, clip, width=0.15\textwidth]{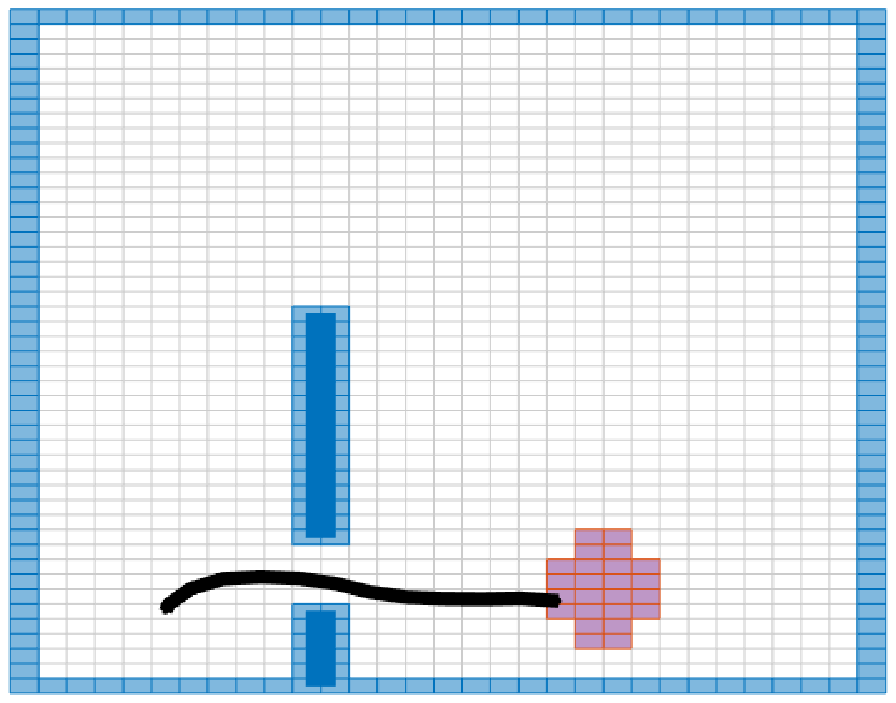}
	\includegraphics[trim={3.3cm 1.8cm 2.8cm 1.8cm}, clip, width=0.15\textwidth]{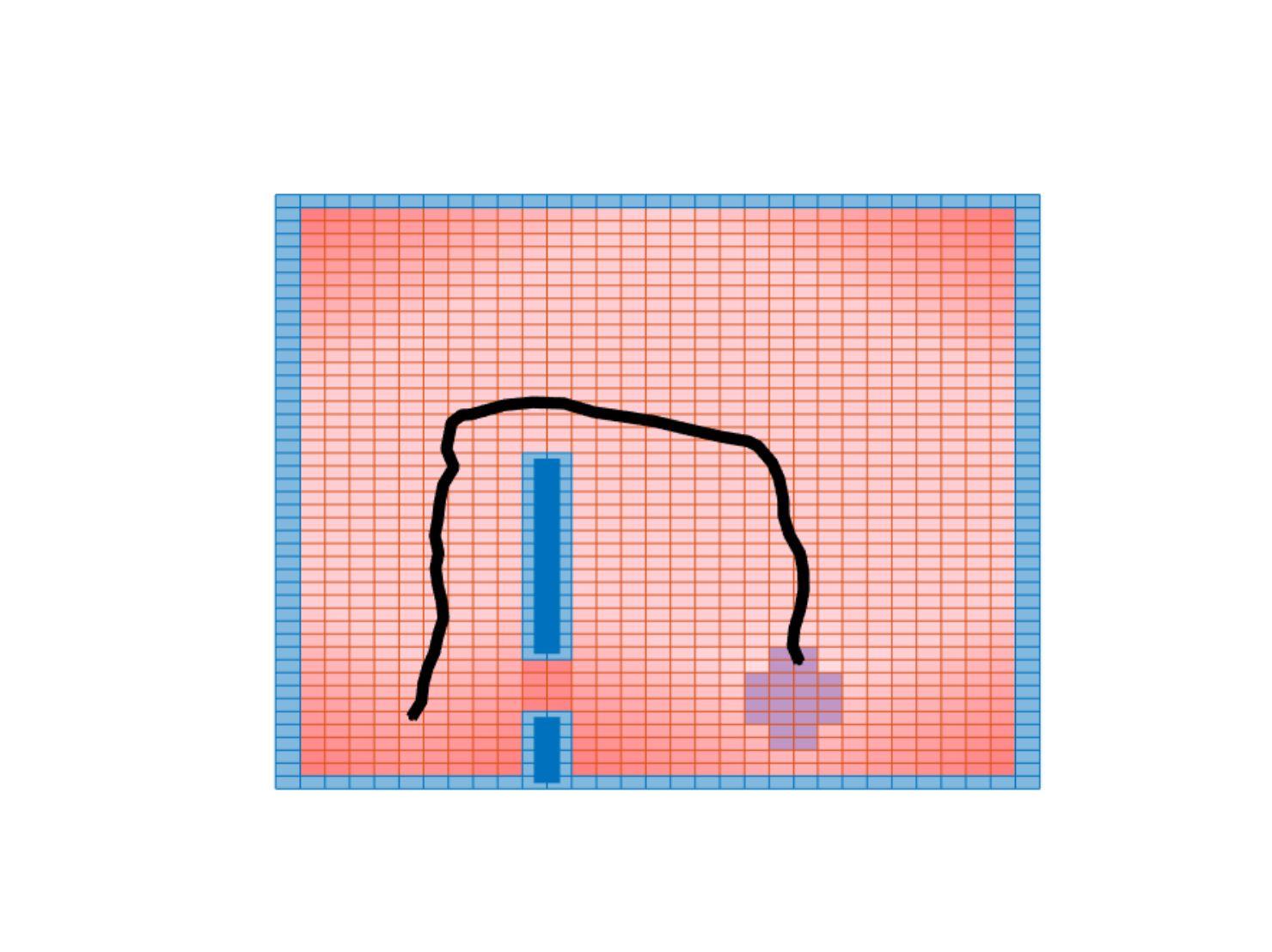}
 	\includegraphics[trim={3.3cm 1.8cm 2.8cm 1.8cm}, clip, width=0.15\textwidth]{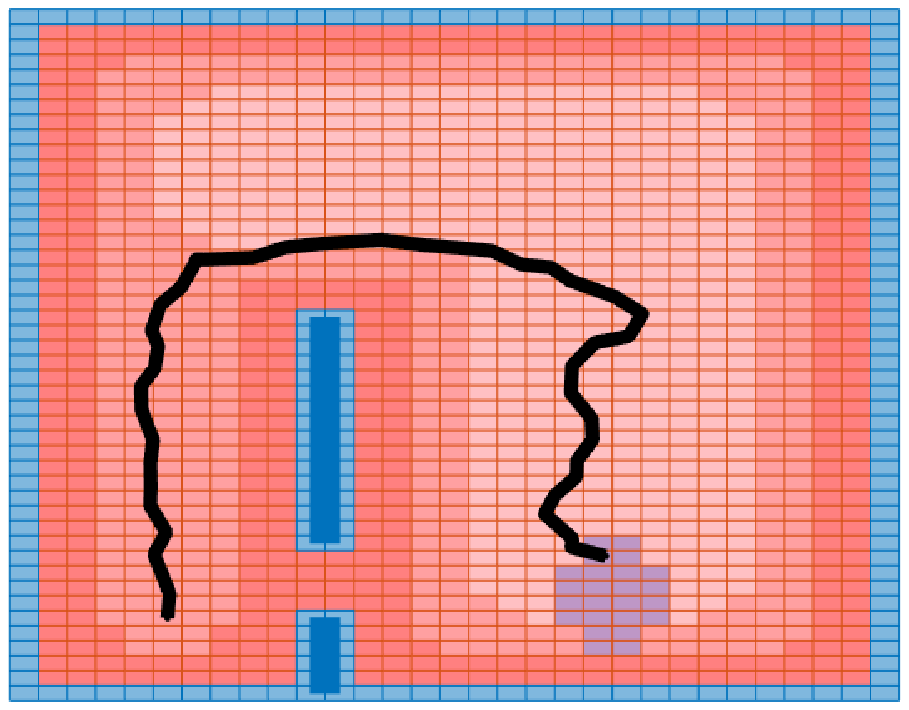}
 	\vspace{-0.3cm}
	 \caption{\textbf{Reach-while-avoid control}: Classical ABCD using SCOTS (left);
	 Resilient ABCD with $d=0.5$ (middle) and $d=2$ (right). Lower resilient states are indicated in darker shade than higher resilient states. Strategies through the wide passage are more resilient.
	 }
	 \label{fig:plots}
\end{figure}

\smallskip
\noindent\textbf{Reach-While-Avoid.}
We first consider a simple \emph{reach-while-avoid} specification for the unicycle robot as depicted in Fig.~\ref{fig:plots}.
In this example, the robot must pass through either a narrow or a wide passage to avoid the obstacles (blue) and reach the target (purple).
We have implemented this synthesis problem as a Parity specification with three colors -- color $2$ assigned to target states and color $1$ assigned to all other states, while color $0$ is kept empty.
To enforce obstacle avoidance, we manipulate the computed abstract transition system to self-loop in obstacle states on all normal and disturbance edges. Hence, whenever the robot visits an obstacle state \emph{once}, it is forced to stay there forever and thereby visits color $1$ infinitely often which violates the specification. Fig.~\ref{fig:plots} depicts the path of the robot to the target; the computed controller additionally ensures that the target is (re-)visited infinitely often, by looping inside it.

Under normal disturbances, both passages are equally safe. Therefore, the original SCOTS algorithm returns a strategy which guides the robot along the shortest path to the target, passing the small passage (Fig.~\ref{fig:plots} (left)).
In contrast, RESCOT synthesizes a controller which enforces the shortest path through the less risky wider passage (Fig.~\ref{fig:plots} (middle and right)) while assuming occasional disturbance spikes with $d=0.5$ (middle) and $d=2$ (right), respectively. 
It can be observed that the smaller $\Whi$ the higher the number of resilience values: 16 for $d=0.5$ (middle) as opposed to 3 for $d=2$ (right) (i.e., the larger the set $\Whi$, the lower the number of high disturbance spikes which pushes the robot from a state in the middle of the work space into an obstacle).

\begin{figure}

\includegraphics[trim={5.5cm 3.4cm 4.8cm 2.5cm}, clip, width=0.23\textwidth]{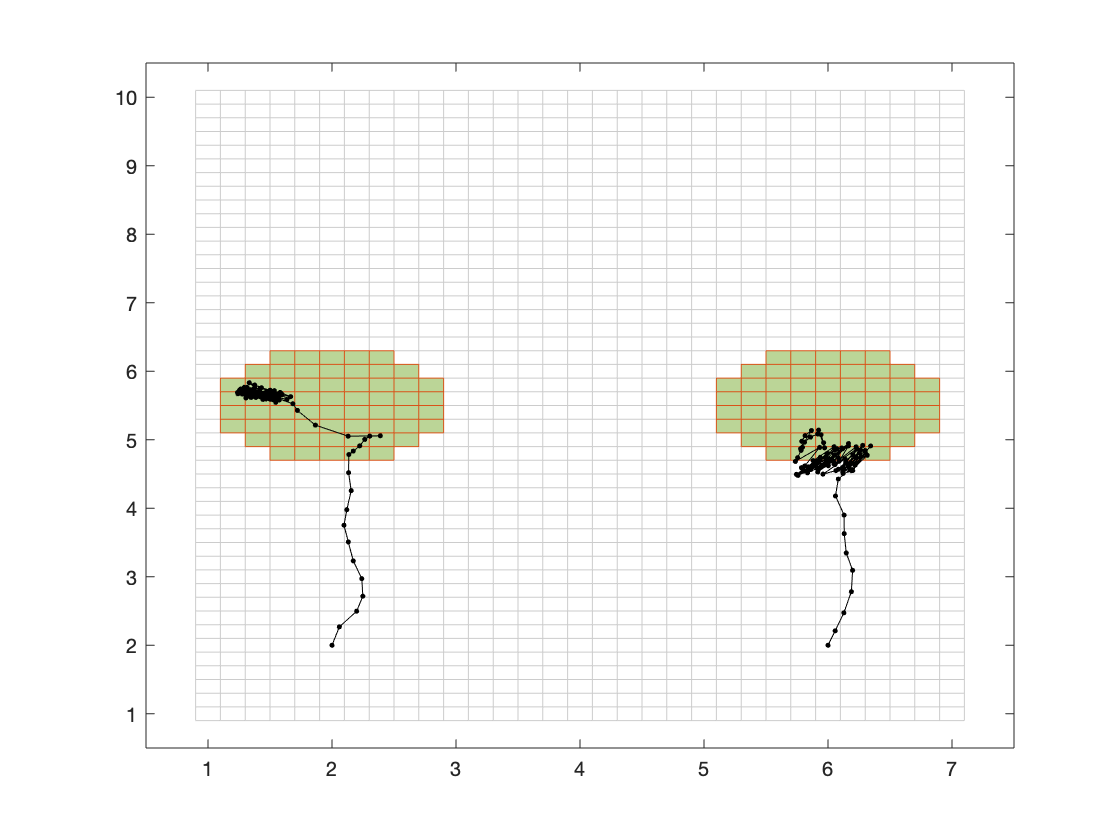}
\includegraphics[trim={5.5cm 3.4cm 4.8cm 2.5cm}, clip, width=0.23\textwidth]{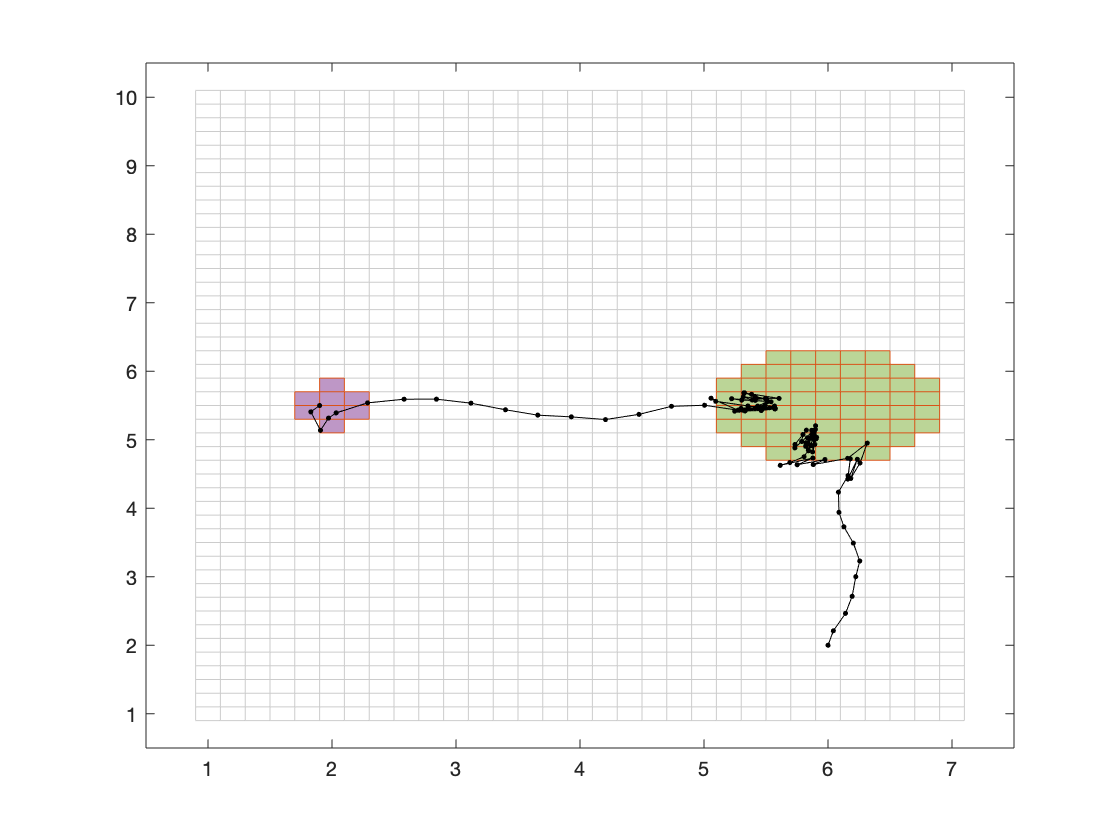}
\vspace{-0.3cm}
\caption{\textbf{Co-Büchi vs. Büchi objectives}: 
	 Fulfilling a Co-Büchi objective for the left and a Büchi objective for the right target. $\omega$ and $\omega+1$ resilient states are indicated in white and green, respectively.
	 }\label{fig:plots_ex2}
	 \vspace{-0.3cm}
\end{figure}

\smallskip
\noindent\textbf{Strategy choices.}
To illustrate the control strategies computed by RESCOT in the presence of 
distinct strategy choices enabled by more expressive specifications,
we consider a $3$-color parity specification for the unicycle robot in a bounded open space with no obstacles.
Here, color $0$ is assigned to the left target ($T_l$), color $2$ is assigned to the right target ($T_r$) and color $1$ is assigned to every other state. 

This specification imposes two possible strategies for the robot: (i)  move to $T_l$ (color 0) and stay inside $T_l$ forever, or (ii) visit at least one state in $T_r$ infinitely often. Hence, the robot can choose between a Co-Büchi specification w.r.t.\ $T_l$ and a Büchi specification w.r.t.\ $T_r$.

The difference of enforced trajectories depending on the Co-Büchi and the Büchi specification is depicted in Fig.~\ref{fig:plots_ex2} (left). We see that the robot does not make an effort to get to the interior of $T_r$, as being pushed out by a disturbance infinitely often (and then moving back in) does not violate the specification. However, the robot makes an effort to stay inside $T_l$; leaving $T_l$ infinitely often would violate the specification. 

For the scenario depicted in Fig.~\ref{fig:plots_ex2} (left) both specifications are satisfyable with resilience $\omega+1$ and the robot chooses the target closest to its starting point. If we decrease the size of the left target (Fig.~\ref{fig:plots_ex2} (right)), only the Büchi specification for $T_r$ is satisfyable with resilience $\omega+1$ and is therefore chosen by the robot from any initial state.

\begin{figure}
 \includegraphics[trim={5.8cm 3.5cm 4.8cm 2.5cm}, clip, width=0.23\textwidth]{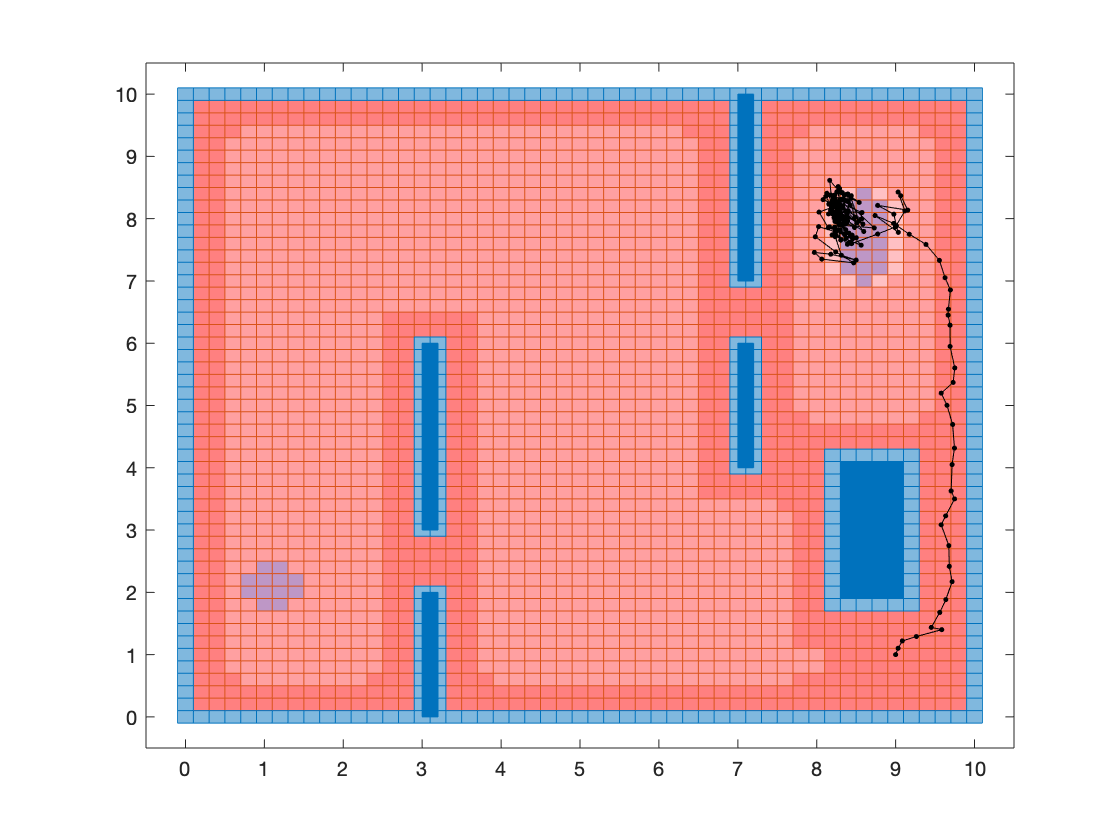}\hfill
\includegraphics[trim={5.8cm 3.5cm 4.8cm 2.5cm}, clip, width=0.23\textwidth]{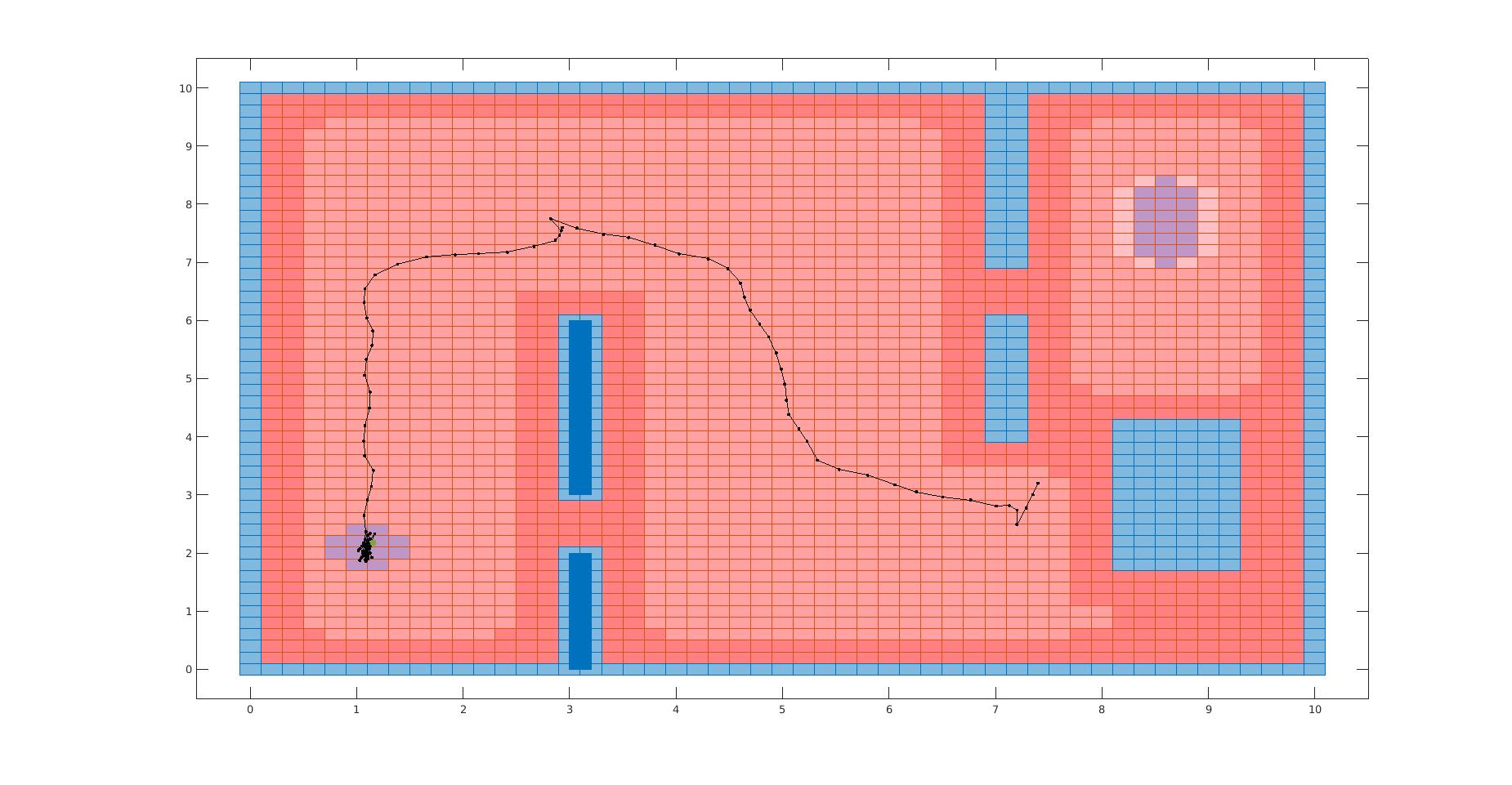}
\vspace{-0.3cm}
	 \caption{Synthesis problem of Fig.~\ref{fig:plots_ex2} under additional obstacle avoidance. All states (including target states) have finite resilience. Here targets are entirely chosen based on the reaching path resilience.
	 }
	 \label{fig:plots_ex3}
\end{figure}

\smallskip
\noindent\textbf{Eventuality properties and finite resilience.}
As a last experiment, we add obstacles to the scenario in Fig.~\ref{fig:plots_ex2} as shown in Fig.~\ref{fig:plots_ex3}. This results in a $4$-color parity game where, as before, color $0$ is assigned to the left target ($T_l$), color $2$ is assigned to the right target ($T_r$) and color $1$ is assigned to every other state. Now we additional assign color $3$ to all obstacles (blue) and again enforce obstacle avoidance by self-loops within those states.

In this case we see that all states, including the states inside the targets, get a finite resilience. This is due to the fact that a finite number of successive disturbances can force the robot to bump into the wall. Choosing to satisfy the Büchi objective (i.e., move to $T_r$) is therefore not more resilient then choosing to satisfy the Co-Büchi objective. In particular, 
the controller avoids passing the risky, dark red passage for moving to $T_r$ and chooses to go to $T_l$ instead, which is further away (see Fig.~\ref{fig:plots_ex3} (right)). However, if it is already inside the risky dark red region, it prefers to move to $T_r$, which is closer.

\begin{table}[h]
\caption{Computation times and controller sizes for the examples discussed in \REFsec{sec:experiments}.}\label{tab:1}
    \centering
    \begin{tabular}{m{2cm}  m{1.5cm}  m{1.5cm} m{1.3cm} }
        \toprule
        & Synthesis time (sec.) & Controller size (KB) & $\#$ colors in $\Phi$ \\
        \midrule
        Fig.~1 (left) & 1453.44 & 90 & N/A  \\
        Fig.~1 (mid) & 26484.4 & 115 & 2 \\
        Fig.~1 (right) & 2948.37 & 90 & 2 \\
        Fig.~2 (left) & 4465 & 98 & 3 \\
        Fig.~2 (right) & 3659 & 61 & 3 \\
        Fig.~3 & 61525.1 & 193 & 4 \\ 
        \bottomrule
    \end{tabular}
\end{table}


\section{Conclusion}
We have developed a novel technique to synthesize resilient controllers for perturbed non-linear dynamical systems w.r.t.\ $\omega$-regular specifications.
Our approach combines two ideas to one effective algorithm: the principle of abstraction-based controller design and a recently developed algorithm for computing optimally resilient strategies in two-player games on graphs.
Our experimental evaluation on various robot motion planning examples has demonstrated that our algorithm indeed produces controllers that are more resilient to disturbance spikes than classical synthesis tools (e.g., SCOTS).

In future work, we are planning to combine RESCOT with our tool Mascot \cite{mascot} to increase computational efficiency and to utilize resilient ABCD within online adaptable ABCD \cite{bai2019incremental} for more realistic applications.

\bibliographystyle{abbrv}
\bibliography{bib}

\end{document}